\documentstyle[aps,multicol,psfig]{revtex}  
\input epsf
\begin{document}
\newcommand{\eqbreak}{
\end{multicols}
\widetext
\noindent
\rule{.48\linewidth}{.1mm}\rule{.1mm}{.1cm}
}
\newcommand{\eqresume}{
\noindent
\rule{.52\linewidth}{.0mm}\rule[-.1cm]{.1mm}{.1cm}\rule{.48\linewidth}{.1mm}
\begin{multicols}{2}
\narrowtext
}
\draft
\title{Universality and its Origins 
at the Amorphous Solidification Transition}
\author{
Weiqun Peng\rlap,$^1$
Horacio E.~Castillo\rlap,$^1$ 
Paul M.~Goldbart$^1$ 
and 
Annette Zippelius$^2$}
\address{
$^1$Department of Physics, 
University of Illinois at Urbana-Champaign, 
1110 West Green Street, 
Urbana, IL 61801-3080, USA}
\address{
$^2$Institut f\"ur Theoretische Physik, 
Universit\"at G\"ottingen, 
D-37073 G\"ottingen, Germany}
\date{\today}
\maketitle
\begin{abstract}
Systems undergoing an equilibrium phase transition from a liquid state
to an amorphous solid state exhibit certain universal characteristics.
Chief among these are the fraction of particles that are randomly localized
and the scaling functions that describe the order parameter and (equivalently) 
the statistical distribution of localization lengths for these localized 
particles.  The purpose of this Paper is to discuss the origins and 
consequences of this universality, and in doing so, three themes are
explored.  First, a replica-Landau-type approach is formulated for the
universality class of systems that are composed of extended objects
connected by permanent random constraints and undergo amorphous
solidification at a critical density of constraints.  This formulation
generalizes the cases of randomly cross-linked and end-linked macromolecular 
systems, discussed previously.  The universal replica free energy is
constructed, in terms of the replica order parameter appropriate to
amorphous solidification, the value of the order parameter is obtained
in the liquid and amorphous solid states, and the chief universal
characteristics are determined.  Second, the theory is reformulated in terms 
of the distribution of local static density fluctuations rather than the replica 
order parameter.  It is shown that a suitable free energy can be constructed, 
depending on the distribution of static density fluctuations, and that this 
formulation yields precisely the same conclusions as the replica approach.
Third, the universal predictions of the theory are compared with the results 
of extensive numerical simulations of randomly cross-linked macromolecular 
systems, due to Barsky and Plischke, and excellent agreement is found.  
\end{abstract}
\pacs{61.43.-j, 82.70.Gg, 64.60.Ak}
%
%
\begin{multicols}{2}
\section{Introduction}
\label{SEC:intro}
During the last decade there has been an ongoing effort to obtain an
ever more detailed understanding of the behavior of randomly
cross-linked macromolecular systems near the vulcanization
transition~\cite{prl_1987,PMGandAZprl,epl,cross}.
This effort has been built from two ingredients: (i)~the Deam-Edwards
formulation of the statistical mechanics of polymer networks~\cite{de};
and (ii)~concepts and techniques employed in the study of spin
glasses~\cite{MPVbook}.  As a result, a detailed mean-field theory for
the vulcanization transition---an example of an amorphous solidification 
transition---has emerged, which makes the following
predictions:
(i)~For densities of cross-links smaller than a certain critical value
(on the order of one cross-link per macromolecule) the system exhibits
a liquid state in which all particles (in the context of macromolecules, 
monomers) are delocalized.  
(ii)~At the critical cross-link density there is a continuous
thermodynamic phase transition to an amorphous solid state, this state
being characterized by the emergence of random static density
fluctuations.
(iii)~In this state, a nonzero fraction of the particles have become
localized around random positions and with random localization lengths
(i.e., r.m.s.~displacements).
(iv)~The fraction of localized particles grows linearly with the excess
cross-link density, as does the characteristic inverse square 
localization length. 
(v)~When scaled by their mean value, the statistical distribution of
localization lengths is universal for all near-critical cross-link
densities, the form of this scaled distribution being uniquely determined
by a certain integro-differential equation. For a detailed review of
these results, see Ref.~\cite{cross}; for an informal discussion, see
Ref.~\cite{REF:apy}.

In the course of the effort to understand the vulcanization transition
for randomly cross-linked macromolecular systems, it has become clear
that one can also employ similar approaches to study randomly 
{\it end-linked\/} macromolecular systems~\cite{end}, and also 
randomly cross-linked {\it manifolds\/} (i.e., higher dimensional
objects)~\cite{manifolds}; in each case, a specific model has been
studied.  For example, in the original case of randomly cross-linked
macromolecular systems, the macromolecules were modeled as flexible,
with a short-ranged excluded-volume interaction, and the cross-links
were imposed at random arc-length locations.  On the other hand, in the
case of {\it end-linked\/} systems, although the excluded-volume
interaction remained the same, the macromolecules were now modeled as
either flexible or stiff, and the random linking was restricted to the
ends of the macromolecules.  Despite the differences between the
unlinked systems and the styles of linking, in all cases
identical critical behavior has been obtained in mean-field theory,
right down to the precise form of the statistical distribution of scaled
localization lengths.

Perhaps even more strikingly, in extensive numerical simulations of 
randomly cross-linked macromolecular systems, Barsky and
Plischke~\cite{REF:SJB_MP_a,REF:SJB_MP_b} have employed an off-lattice 
model of macromolecules interacting via a Lennard-Jones potential.  Yet 
again, an essentially identical picture has emerged for the transition 
to and properties of the amorphous solid state, despite the substantial
differences between physical ingredients incorporated in the simulation
and in the analytical theory.

In the light of these observations, it is reasonable to ask whether one
can find a common theoretical formulation of the amorphous
solidification transition (of which the vulcanization transition is a
prime example) that brings to the fore the emergent collective
properties of all these systems that are model-independent, and
therefore provide useful predictions for a broad class of experimentally
realizable systems.  The purpose of this Paper is to explain how this
can be done.  In fact, we approach the issue in two distinct (but
related) ways, in terms of a replica order parameter and in terms of the
distribution of random static density fluctuations, either of which
can be invoked to characterize the emergent amorphous solid state.

The outline of this Paper is as follows. 
In Sec.~\ref{SEC:MFT_replicas} we construct the universal replica Landau
free energy of the amorphous solidification transition. In doing this,
we review the replica order parameter for the amorphous solid state and
discuss the constraints imposed on the replica Landau free energy by
(a)~symmetry considerations, (b)~the smallness of the fraction of
particles that are localized near the transition, and (c)~the weakness
of the localization near the transition.
In Sec.~\ref{SEC:MFT_solve} we invoke a physical hypothesis to 
solve the stationarity condition for the replica order parameter, 
thereby obtaining a mean-field theory of the transition.  We exhibit the 
universal properties of this solution and, in particular, the scaling 
behavior of certain central physical quantities. 
In Sec.~\ref{SEC:fluctuations} we describe an alternative approach to
the amorphous solidification transition, in which we construct and
analyze the Landau free energy expressed in terms of the distribution of
static density fluctuations. Although we shall invoke the replica
approach in the construction of this Landau free energy, its ultimate
form does not refer to replicas. As we show, however, the physical
content of this Landau theory is identical to that of the replica Landau
theory addressed in Secs.~\ref{SEC:MFT_replicas} and \ref{SEC:MFT_solve}.
In Sec.~\ref{SEC:simulations} we exhibit the predicted
universality by examining the results of extensive numerical simulations
of randomly cross-linked macromolecular networks, due to Barsky and
Plischke. 
In Sec.~\ref{SEC:conclusions} we give some concluding remarks. 
\section{Universal replica free energy for 
the amorphous solidification transition}
\label{SEC:MFT_replicas}
We are concerned, then, with systems of extended objects, such as
macromolecules, that undergo a transition to a state characterized by
the presence of random static fluctuations in the particle-density when
subjected to a sufficient density of permanent random constraints (the
character and statistics of which constraints preserve translational and
rotational invariance).  In such states, translational and rotational
symmetry are spontaneously broken, but in a way that is hidden at the
macroscopic level.  We focus on the long wavelength physics in the
vicinity of this transition.
 
In the spirit of the standard Landau approach, we envisage that the
replica technique has been invoked to incorporate the consequences of
the permanent random constraints, and propose a phenomenological
mean-field replica free energy, the $n\to 0$ limit of which gives the
disorder-averaged free energy, in the form of a power series in the
replica order parameter.  We invoke symmetry arguments, along with the
recognition that near to the transition the fraction of particles that
are localized is small and their localization is weak. The control
parameter $\epsilon$ is proportional to the amount by which the
constraint density exceeds its value at the transition.  As we shall
see, the stationarity condition for this general, symmetry-inspired
Landau free energy is satisfied by precisely the order-parameter
hypothesis that exactly solves the stationarity conditions derived from
semi-microscopic models of cross-linked and end-linked 
macromolecules.  From the properties of this solution we recover the 
primary features of the liquid-amorphous solid transition.

In a system characterized by static random density-fluctuations, one
might na{\"\i}vely be inclined to use the particle-density as the order
parameter.  However, the disorder-averaged particle density cannot
detect the transition between the liquid and the amorphous solid states,
because it is uniform (and has the same value) in both states:  a
subtler order parameter is needed.  As shown earlier, for the specific
cases of randomly cross-linked~\cite{prl_1987,epl,cross} and
end-linked~\cite{end} macromolecular systems, the appropriate order
parameter is instead:
\begin{equation}
\Omega_{{\bf k}^{1},{\bf k}^{2},\cdots,{\bf k}^{g}}
\equiv
\left[
\frac{1}{N}\sum_{i=1}^{N}
\langle e^{i{\bf k}^{1}\cdot{\bf c}_{i}} \rangle_{\chi}
\langle e^{i{\bf k}^{2}\cdot{\bf c}_{i}} \rangle_{\chi}
\cdots
\langle e^{i{\bf k}^{g}\cdot{\bf c}_{i}} \rangle_{\chi}
\right], 
\label{EQ:opDefinition}
\end{equation}
where 
$N$ is the total number of particles, 
${\bf c}_{i}$ (with $i=1,\ldots, N$) is the position of particle $i$, 
the wave-vectors ${\bf k}^{1},{\bf k}^{2},\cdots,{\bf k}^{g}$ are arbitrary, 
$\langle\cdots\rangle_{\chi}$ denotes a thermal average for a particular 
realization $\chi$ of the disorder, and $\left[\cdots\right]$ represents  
averaging over the disorder.

We make the Deam-Edwards assumption~\cite{de} that the statistics of the
disorder is determined by the instantaneous correlations of the
unconstrained system.  (It is as if one took a snapshot of the system
and, with some nonzero probability, added constraints only at those
locations where two particles are in near contact.)\thinspace\  This
assumption leads to the need to work with the $n\to 0$ limit of systems
of $n+1$, as opposed to $n$, replicas. The additional replica, labeled
by $\alpha=0$, represents the degrees of freedom of the original system
before adding the constraints, or, equivalently, describes the
constraint distribution.

We assume, for the most part, that the free energy is invariant under
the group $S_{n+1}$ of permutations of the ${n+1}$ replicas.  In this
replica formalism, the replica order parameter turns out to be
\begin{equation}
\Omega_{\hat{k}}\equiv
\Big\langle
\frac{1}{N}\sum_{i=1}^{N}
\exp\big(i{\hat{k}}\cdot{\hat c}_{i}\big)
\Big\rangle_{n+1}^{\rm P}.
\label{EQ:ReplicaOrder}
\end{equation}
Here, hatted vectors denote replicated collections of vectors, viz., 
${\hat{v}}\equiv\{{\bf v}^{0},{\bf v}^{1},\cdots,{\bf v}^{n}\}$,
their scalar product being 
${\hat{v}}\cdot{\hat{w}}\equiv
\sum_{\alpha=0}^{n}{\bf v}^{\alpha}\cdot{\bf w}^{\alpha}$, 
and $\langle\cdots\rangle_{n+1}^{\rm P}$ denotes an average for an
effective pure (i.e., disorder-free) system of $n+1$ coupled replicas of
the original system. We use the terms {\it one-replica sector\/} and
{\it higher-replica sector\/} to refer to replicated vectors with,
respectively, exactly one and more than one replica $\alpha$  for which
the corresponding vector ${\bf k}^{\alpha}$ is nonzero.  In particular,
the order parameter in the one-replica sector reduces to the
disorder-averaged mean particle-density, and plays only a minor role in
what follows.  The appearance of $n+1$ replicas in the order parameter
allows one to probe the correlations between the density fluctuations in
the constrained system and the density fluctuations in the unconstrained
one.

We first study the transformation properties of the order parameter
under translations and rotations, and then make use of the resulting
information to determine the possible terms appearing in the replica
free energy.  Under independent translations of all the replicas, i.e., 
${\bf c}_{i}^{\alpha}\rightarrow{\bf c}_{i}^{\alpha}+{\bf a}^{\alpha}$,
the replica order parameter, Eq.~(\ref{EQ:ReplicaOrder}), transforms as 
\begin{equation}
\Omega_{\hat{k}}\rightarrow
\Omega^{\prime}_{\hat{k}}= 
{\rm e}^{i\sum_{\alpha=0}^{n}{\bf k}^{\alpha}\cdot{\bf a}^{\alpha}} \Omega_{\hat{k}}.
\label{EQ:Omega_trans}
\end{equation}
Under independent rotations of the replicas, defined by
${\hat{R}}{\hat{v}}\equiv 
\{R^0{\bf v}^{0},\cdots,R^{n}{\bf v}^{n}\}$ and
${\bf c}_{i}^{\alpha} \rightarrow R^{\alpha}{\bf c}_{i}^{\alpha}$, 
the order parameter transforms as
\begin{equation}
\Omega_{\hat{k}}\rightarrow
\Omega^{\prime}_{\hat{k}}=
\Omega_{\hat{R}^{-1}\hat{k}}.
\label{EQ:Omega_rot}
\end{equation}

As discussed in detail in~\cite{cross}, because we are concerned with
the transition between liquid and amorphous solid states, both of which
have uniform macroscopic density, the one-replica sector order parameter
is zero on both sides of the transition. This means that the sought free
energy can be expressed in terms of contributions referring to the
higher-replica sector order parameter, alone.

We express the free energy as an expansion in (integral) powers of the
replica order parameter $\Omega_{\hat{k}}$,  retaining the two lowest
possible powers of $\Omega_{\hat{k}}$, which in this case are the square
and the cube.  We consider the case in which no external potential
couples to the order parameter.  Hence, there is no term linear in the
order parameter. We make explicit use of translational symmetry,
Eq.~(\ref{EQ:Omega_trans}), and thus obtain the following expression for
the replica free energy (per particle per space dimension) 
${\cal F}_{n}(\{\Omega_{\hat{k}}\})$ \cite{notation}:
\begin{eqnarray}
nd&{\cal F}_{n}&\big(\{\Omega_{\hat{k}}\}\big) = 
{\overline{\sum}}_{\hat{k}} g_{2}( {\hat{k}}) | \Omega_{\hat{k}} |^{2} 
\nonumber \\
& - & {\overline{\sum}}_{
{\hat{k}_1} {\hat{k}_2} {\hat{k}_3}}
g_{3}\big(\hat{k}_1,\hat{k}_2,\hat{k}_3\big)
\Omega_{\hat{k}_1} 
\Omega_{\hat{k}_2} 
\Omega_{\hat{k}_3} 
\delta_{{\hat{k}_1}+{\hat{k}_2}+{\hat{k}_3},{\hat{0}}}\,.
\label{EQ:LG_general}
\end{eqnarray}
Here, the symbol ${\overline{\sum}}$ denotes a sum over replicated
vectors $\hat{k}$ lying in the higher-replica sector, and we have made
explicit the fact that the right hand side is linear in $n$ (in the 
$n\rightarrow 0$ limit) by factoring $n$ on the left hand side.  
In a microscopic approach, the functions $g_{2}(\hat{k})$ and
$g_{3}\big(\hat{k}_1,\hat{k}_2,\hat{k}_3\big)$ would be obtained in
terms of the control parameter $\epsilon$, together with density
correlators of an uncross-linked liquid having interactions
renormalized by the cross-linking.  Here, however, we will ignore the
microscopic origins of $g_{2}$ and $g_{3}$, and instead use symmetry
considerations and a long-wavelength expansion to determine only their 
general forms.  In the saddle-point approximation, then, the 
disorder-averaged free energy $f$ (per particle and space dimension) 
is given by ~\cite{MPVbook}:
\begin{equation}
f=\lim_{n\rightarrow 0}\min_{\{\Omega_{\hat{k}}\}} 
{\cal F}_{n}\big(\{\Omega_{\hat{k}}\}\big). 
\label{EQ:physical_f}
\end{equation}  

Bearing in mind the physical notion that near the transition 
any localization should occur only on long length-scales, we 
examine the long wavelength limit by also performing a low-order 
gradient expansion.  In the term quadratic in the order parameter 
we keep only the leading and next-to-leading order
terms in $\hat{k}$; 
in the cubic term in the order parameter we keep only 
the leading term in $ {\hat{k}}$. 
Thus, the function $g_{3}$ in Eq.~(\ref{EQ:LG_general}) is
replaced by a constant and the function $g_{2}$ is expanded to
quadratic order in $\hat{k}$. By analyticity and 
rotational invariance, $g_{2}$ can only depend on 
$\big\{{\bf k}^{0},\ldots,{\bf k}^{n}\big\}$ via 
$\big\{\vert{\bf k}^{0}\vert^2,\ldots,\vert{\bf k}^{n}\vert^2\big\}$, 
and, in particular, terms linear in $\hat{k}$ are excluded. In addition,
by the permutation symmetry among the replicas, each term $\vert{\bf
k}^{\alpha}\vert^2$ must enter the expression for $g_{2}$ with a common
prefactor, so that the dependence is in fact on $\hat{k}^{2}$. 
Thus, the replica free energy for long-wavelength 
density fluctuations has the general form:
\begin{eqnarray}
&&nd{\cal F}_{n}\big(\{\Omega_{\hat{k}}\}\big) = 
{\overline{\sum}}_{\hat{k}}
\Big(-a\epsilon+\frac{b}{2}|\hat{k}|^2\Big)
\big\vert\Omega_{\hat{k}}\big\vert^{2}
\nonumber\\
&&\qquad\qquad
-c\,{\overline{\sum}}_{{\hat{k}_1}{\hat{k}_2}{\hat{k}_3}}
\Omega_{\hat{k}_1}\,
\Omega_{\hat{k}_2}\,
\Omega_{\hat{k}_3}\,
\delta_{{\hat{k}_1}+{\hat{k}_2}+{\hat{k}_3}, {\hat{0}}}\,.
\label{EQ:LG_longwave}
\end{eqnarray}To streamline the presentation, we take advantage of the 
freedom to rescale ${\cal F}_{n}$, $\epsilon$ and $\hat{k}$, thus 
setting to unity the parameters $a$, $b$ and $c$.  Thus, the free 
energy becomes
\begin{eqnarray}
&&nd{\cal F}_{n}\big(\{\Omega_{\hat{k}}\}\big) = 
{\overline{\sum}}_{\hat{k}}
\Big(-\epsilon+\frac{|\hat{k}|^2}{2}\Big)
\big\vert\Omega_{\hat{k}}\big\vert^{2}
\nonumber\\
&&\qquad\qquad
-\,{\overline{\sum}}_{{\hat{k}_1}{\hat{k}_2}{\hat{k}_3}}
\Omega_{\hat{k}_1}\,
\Omega_{\hat{k}_2}\,
\Omega_{\hat{k}_3}\,
\delta_{{\hat{k}_1}+{\hat{k}_2}+{\hat{k}_3}, {\hat{0}}}\,.
\label{EQ:LG_rescale}
\end{eqnarray}

By taking the first variation with 
respect to $\Omega_{-\hat{k}}$ we obtain the stationarity condition 
for the replica order parameter:
\begin{eqnarray}
0&=&nd\frac{\delta{\cal F}_{n}}{\delta \Omega_{-\hat{k}}} 
\nonumber\\&=& 
2\Big(-\epsilon+\frac{\vert{\hat{k}}\vert^2}{2}\Big)
\Omega_{\hat{k}} 
-3\overline{\sum\limits_{\hat{k}_1\hat{k}_2}}
\Omega_{\hat{k}_1}\,
\Omega_{\hat{k}_2}\,
\delta_{{\hat{k}_1}+{\hat{k}_2},{\hat{k}}}\,.
\label{EQ:LG_saddle}
\end{eqnarray}
This self-consistency condition applies for all values of $\hat{k}$ 
lying in the higher-replica sector.  
\section{Universal properties of the order parameter 
in the amorphous solid state}
\label{SEC:MFT_solve}
Generalizing what was done for cross-linked and end-linked
macromolecular systems, we hypothesize that the particles have a
probability $q$ of being localized (also called 
the ``gel fraction'' in the context of vulcanization) and
$1-q$ of being delocalized, and that the
localized particles are characterized by a probability distribution
$2\xi^{-3}p(\xi^{-2})$ for their localization lengths $\xi$.  Such a
characterization weaves in the physical notion that amorphous systems
should show a spectrum of possibilities for the behavior of their
constituents, and adopts the perspective that it is this spectrum that
one should aim to calculate.  This hypothesis translates into the
following expression for the replica order parameter~\cite{epl,cross}:
\begin{equation}
\Omega_{\hat{k}}= 
(1-q)\,\delta_{ {\hat{k}},\hat{0}} 
+q\,\delta^{(d)}_{\tilde{\bf k},{\bf 0}}\,
\int_{0}^{\infty}\!\!d\tau\,p(\tau)\,
{\rm e}^{-\hat{k}^{2}/2\tau},
\label{EQ:ord_par_hyp}
\end{equation}
where we have used the notation  
$\tilde{\bf k}\equiv\sum_{\alpha=0}^{n}{\bf k}^\alpha$.  The first term
on the right hand side term represents  delocalized particles, and is
invariant under independent translations of each replica
(cf.~Eq.~\ref{EQ:Omega_trans}).  In more physical terms, this
corresponds to the fact that not only the average particle density but
the individual particle densities are translationally invariant.  The
second term represents particles that are localized, and is only
invariant under common translations of the replicas (i.e., translations
in which ${\bf a}^{\alpha}= {\bf a}$ for all $\alpha$). This corresponds
to the fact that the individual particle density for localized particles
is not translationally invariant, so that translational invariance is
broken microscopically, but the average density remains translationally
invariant, i.e.~the system still is macroscopically translationally
invariant (MTI).

By inserting the hypothesis~(\ref{EQ:ord_par_hyp}) into the stationarity
condition~(\ref{EQ:LG_saddle}), and taking the $n\rightarrow 0$ limit,
we obtain
\begin{eqnarray}
&&0=\delta^{(d)}_{\tilde{\bf k},{\bf 0}}
\left\{ 
2\left(3q^{2}-\epsilon q+q
{{\hat{k}}^2}/{2}
\right)
\int_{0}^{\infty}\!\!d\tau\,p(\tau)\,
\,{\rm e}^{-\hat{k}^{2}/2\tau}
\right. 
\nonumber\\
&&\,\,-\left.
3q^{2}\!
\int_{0}^{\infty}\!\!\!d\tau_1\,p(\tau_1) 
\int_{0}^{\infty}\!\!\!d\tau_2\,p(\tau_2)
\,{\rm e}^{-\hat{k}^{2}/2(\tau_1+\tau_2)}\right\}.
\nonumber\\
\label{EQ:int_saddle}
\end{eqnarray}
In the limit ${\hat{k}}^{2}\rightarrow 0$, the equation reduces to 
a condition for the localized fraction $q$, viz.,
\begin{equation}
0=-2q\epsilon+3q^2. 
\label{EQ:q_eqn}
\end{equation}
For negative or zero $\epsilon$, corresponding to a constraint density
less than or equal to its critical value, the only physical solution is
$q = 0$, corresponding to the liquid state. In this state, all particles
are delocalized. For positive $\epsilon$, corresponding to a constraint
density in excess of the critical value, there are two solutions. One,
unstable, is the continuation of the liquid state $q=0$; the other,
stable, corresponds to a nonzero fraction,
\begin{equation}
q=\frac{2}{3}\epsilon
\label{EQ:q_soln}
\end{equation}
being localized. We identify this second
state as the amorphous solid state. From the dependence of the localized
fraction $q$ on the control parameter $\epsilon$ and the form of the
order parameter Eq.~(\ref{EQ:ord_par_hyp}) we conclude that there is a
continuous phase transition between the liquid and the amorphous solid
states at $\epsilon=0$, with localized fraction exponent $\beta=1$
(i.e., the classical exponent \cite{REF:classical}).  
It is worth mentioning that microscopic
approaches go beyond this linear behavior near the transition, yielding 
a transcendental equation for $q(\epsilon)$, valid for all values of the 
control parameter $\epsilon$; see Ref.~\cite{epl,cross}. From 
Eq.~\ref{EQ:LG_rescale} it is evident that the liquid
state is locally stable (unstable) for negative (positive) $\epsilon$:
the eigenvalues of the resulting quadratic form are given by
$\lambda(\hat{k})=-\epsilon+{\hat{k}^2}/2$. 

Now concentrating on the amorphous solid state, by inserting the
value of the localized fraction into Eq.~(\ref{EQ:int_saddle}),
we obtain the following integro-differential equation for the
probability distribution for the localization lengths:
\begin{equation}
\frac{\tau^2}{2}\frac{dp}{d\tau}=
\Big(\frac{\epsilon}{2}-\tau\Big)p(\tau)
-\frac{\epsilon}{2}\int_{0}^{\tau}\!\!d\tau_1\,
 p(\tau_1)\,p(\tau-\tau_1).
\label{EQ:scp}   
\end{equation}
All parameters can be seen to play an elementary role in this equation
by expressing $p(\tau)$ in terms of a scaling function:
\begin{equation}
p(\tau)=({2}/{\epsilon})\,\pi(\theta);
\qquad
\tau=({\epsilon}/{2})\,\theta.
\label{EQ:scale}
\end{equation} 
Thus, the universal function $\pi(\theta)$ satisfies 
\begin{equation}
\frac{\theta^{2}}{2} \frac{d\pi}{d\theta}
= (1-\theta)\,\pi(\theta)-
\int_{0}^{\theta} d\theta^{\prime}
\pi(\theta^{\prime})\pi(\theta-\theta^{\prime}). 
\label{EQ:scpieq}
\end{equation}
Solving this equation, together with the normalization condition
$1=\int\nolimits_{0}^{\infty}d\theta\,\pi(\theta)$, one finds the 
scaling function shown in Refs.~\cite{epl,cross}.  The function
$\pi(\theta)$ has a peak at $\theta\simeq 1$ of width of order unity,
and decays rapidly both as $\theta\to 0$ and $\theta\to\infty$.  By
combining these features of $\pi(\theta)$ with the scaling
transformation~(\ref{EQ:scale})  we conclude that the typical
localization length scales as $\epsilon^{-1/2}$ near the transition. 
The order parameter also has a scaling form near the transition:
\begin{eqnarray}
\Omega_{\hat{k}}
&=&
\Big(1-\big(2\epsilon/3\big)\Big)
\delta_{\hat{k},\hat{0}}+
\Big(2\epsilon/3\Big)
\delta^{(d)}_{\tilde{\bf k},{\bf 0}}\,
\omega\Big(\sqrt{2\hat{k}^{2}/\epsilon}\Big),
\nonumber\\
\omega(k)
&=&
\int_{0}^{\infty}d\theta\,\pi(\theta)
{\rm e}^{-k^{2}/2\theta}\,.
\label{EQ:ord_par_scale}
\end{eqnarray}Equation~(\ref{EQ:scpieq}) 
and the normalization condition on $\pi(\theta)$ are precisely the
conditions that determine the scaling function for the cross-linked and
end-linked cases; they are discussed extensively, together with the
properties of the resulting distribution of localization lengths and
order parameter, in Refs.~\cite{epl,cross,end}.

As discussed in this section, the localized fraction $q(\epsilon)$ and
the scaled distribution of inverse square localization lengths
$\pi(\theta)$ are universal near the transition.  We now discuss this 
issue in more detail.  

First, let us focus at the mean-field level.  Recall the mean-field theory 
of ferromagnetism~\cite{REF:Ma} and, in particular, the exponent $\beta$, 
which characterizes the vanishing of the magnetization density order 
parameter (from the ferromagnetic state) as a function of the temperature at 
zero applied magnetic field. The exponent $\beta$ takes on the value of 1/2,
regardless of the details of the mean-field theory used to compute it.
The functions $q(\epsilon)$, $\pi(\theta)$ and $\omega(k)$ are universal 
in the same sense. The case of $q(\epsilon)$ is on essentially the same,
standard, footing as that of the magnetization density. What is not standard,
however, is that describing the (equilibrium) order parameter 
is a universal scaling {\it function\/}, $\omega(k)$ [or, equivalently, 
$\pi(\theta)$] that is not a simple power law. This feature arises because 
the usual presence of fields carrying {\it internal\/} indices, such 
as cartesian vector indices in the case of ferromagnetism, is replaced 
here by the external continuous replicated wave vector variable $\hat{k}$. 
There are two facets to this universal scaling behavior of the order 
parameter.  First, for systems differing in their microscopic details and 
their constraint densities there is the possibility of collapsing the 
distribution of localization lengths on to a single function, solely by 
rescaling the independent variable.  Second, there is a definite prediction 
for the dependence of this rescaling on the control parameter $\epsilon$. 

Now, going beyond the mean-field level, in the context of vulcanization 
de~Gennes~\cite{DeGennes} has shown that the width of the critical
region, in which fluctuations dominate and mean-field theory fails, 
vanishes in the limit of very long macromolecules in space-dimension $d=3$ 
or higher.  Thus, one may anticipate that for extended objects the 
mean-field theory discussed here will be valid, except in an 
exceedingly narrow region around the transition.  Nevertheless, if---as 
is usually the case---the effective hamiltonian governing the fluctuations 
is the Landau free energy then the universality discussed here is expected 
to extend, {\it mutatis mutandis\/}, into the critical regime. 

That the amorphous solid state given by
Eq.~(\ref{EQ:ord_par_scale}) is stable with respect to small
perturbations (i.e., is locally stable) can be shown by detailed analysis.
Moreover, as we shall see in
Sec.~\ref{SEC:simulations}, it yields predictions that are in
excellent agreement with subsequent computer simulations.  However,
there is, in principle, no guarantee that this state is globally stable
(i.e., that no states with lower free energy exist).

Up to this point we have assumed that the free energy is
invariant under interchange of all replicas, including the one 
representing the constraint distribution ($\alpha=0$), with any of the
remaining $n$, i.e.~that the free energy is symmetric
under the group $S_{n+1}$ of permutations of all $n+1$ replicas.  This
need not be the case, in general, as the system can be changed, (e.g.,
by changing the temperature) after the constraints have been imposed. In
this latter case, the free energy retains the usual $S_n$ symmetry under
permutations of replicas $\alpha=1,\ldots,n$. The argument we have
developed can be reproduced for this more general case with only a minor
change: in the free energy, we can no longer invoke $S_{n+1}$ symmetry
to argue that all of the $\vert{\bf k^{\alpha}}\vert^2$ must enter the
expression for $g_{2}$ with a common prefactor. Instead, we only have
permutation symmetry among replicas $\alpha=1,\ldots,n$ and, therefore,
the prefactor $b$ for all of these replicas must be the same, but now
the prefactor $b_0$ for replica $\alpha=0$ can be different.  This
amounts to making the replacement
\begin{equation}
\hat{k}^2\rightarrow\bar{k}^2 
\equiv
b_0 b^{-1}\big\vert{\bf k^{0}}\big\vert^2 
+\sum\nolimits_{\alpha=1}^{n}\big\vert{\bf k^{\alpha}}\big\vert^2
\end{equation}
in the free energy.  Both the rest of the derivation and the results are
unchanged, except that $\hat{k}^2$ needs to be replaced by $\bar{k}^2$,
throughout. We mention, in passing, that no saddle points exhibiting the
spontaneous breaking of replica permutation symmetry have been found, to
date, either for systems with $S_{n+1}$ or $S_{n}$ symmetry.
\section{Free energy in terms of the 
distribution of static density fluctuations}
\label{SEC:fluctuations}
The aim of this section is to construct an expression for the
disorder-averaged Landau free energy for the amorphous
solidification transition, ${\cal F}$, in terms of the distribution of
static density fluctuations.  We present this approach as an alternative
to the strategy of constructing a replica free energy ${\cal F}_{n}$ in
terms of the replica order parameter $\Omega$.  In the familiar way, the
equilibrium state will be determined via a variational principle:
$\delta{\cal F}=0$ and $\delta^{2}{\cal F}>0$.  What may be less
familiar, however, is that in the present setting the 
{\it independent\/} variables for the variation (i.e. the distribution of
static density fluctuations) themselves constitute a
functional. 

Our aim, then, is to work not with the replica order parameter
$\Omega_{\hat k}$, but instead with the disorder-averaged probability
density functional for the random static density
fluctuations~\cite{sdf,cross}, ${\cal N}\big(\{\rho_{\bf k}\}\big)$,
which is defined via
\begin{equation}
{\cal N}\big(\{\rho_{\bf k}\}\big)
\equiv
\left[
\frac{1}{N}{\sum_{i=1}^{N}}
{\prod_{{\bf k}}}
{\delta_{\rm c}}\Big(
\rho_{\bf k}-
\langle\exp\big(i{\bf k}\cdot{\bf c}_{i}\big)\rangle_{\chi}\Big)
\right].
\label{EQ:stat_den_dist}
\end{equation}
Here, $\prod\nolimits_{{\bf k}}$ denotes the product over all
$d$-vectors ${\bf k}$, and the Dirac $\delta$-function of complex
argument ${\delta_{\rm c}}(z)$ is defined by 
${\delta_{\rm c}}(z)\equiv
\delta({{\rm Re\,}} z)\,\delta({{\rm Im\,}} z)$, 
where ${{\rm Re\,}} z$ and ${{\rm Im\,}} z$ respectively denote the real
and imaginary parts of the complex number $z$.   From the definition of
${\cal N}\big(\{\rho_{\bf k}\}\big)$, we see that ${\rho_{-{\bf k}}}=
{\rho_{\bf k}^{\ast}}$ and $\rho_{\bf 0}=1$. Thus we can take as
independent variables $\rho_{\bf k}$ for all $d$-vectors ${\bf k}$ in
the half-space given by the condition ${\bf k}\cdot{\bf n}>0$ for a
suitable unit $d$-vector ${\bf n}$.  In addition, 
${\cal N}\big(\{\rho_{\bf k}\}\big)$ obeys the normalization condition
\begin{equation}
\int{\cal D}\rho\,{\cal N}\big(\{\rho_{\bf k}\}\big)=1.
\label{EQ:N_norm}
\end{equation}
It is straightforward to check that, for any particular positive integer
$g$, the replica order parameter $\Omega_{\hat k}$ is a $g^{\rm th}$
moment of ${\cal N}\big(\{\rho_{\bf k}\}\big)$:
\begin{equation}
\int{\cal D} \rho\,\,
{\cal N}\big(\{\rho_{\bf k}\}\big)\,
\rho_{{\bf k}^{1}}\,
\rho_{{\bf k}^{2}}\,
\cdots\,
\rho_{{\bf k}^{g}}
=\Omega_{{\bf k}^{1},{\bf k}^{2},\ldots,{\bf k}^{g}}, 
\label{EQ:moment_stat_den}
\end{equation} 
We use ${\cal D}\rho$ to denote the measure 
${\prod_{{\bf k}}}
d\,{{\rm Re\,}}\rho_{\bf k}\, 
d\,{{\rm Im\,}}\rho_{\bf k}$.

The merit of the distribution functional 
${\cal N}\big(\{\rho_{\bf k}\}\big)$ over the replica order parameter 
$\Omega_{\hat k}$ is that, as we shall soon see, it allows us to
formulate a Landau free energy for the amorphous solidification 
transition, depending on ${\cal N}\big(\{\rho_{\bf k}\}\big)$, in which
replicated quantities do not appear, while maintaining the physical
content of the theory.  At the present time, this approach is not 
truly independent of the replica approach, in the
following sense: we employ the replica approach to derive the free
energy, Eq.~(\ref{EQ:LG_rescale}), 
and only then do we transform from the language of order
parameters to the language of the distribution of static density
fluctuations.  We are not yet in possession of either an analytical
scheme or a set of physical arguments that would allow us to construct the
Landau free energy directly.  Nevertheless, we are able, by
this indirect method, to propose a (replica-free) free energy, and also 
to hypothesize (and verify the correctness of) a stationary value of 
${\cal N}\big(\{\rho_{\bf k}\}\big)$. It would, however, be very 
attractive to find a scheme that would allow us to eschew the replica 
approach and work with the distribution of static density fluctuations 
from the outset.

To proceed, we take the replica Landau free energy ${\cal F}_{n}$, 
Eq.~(\ref{EQ:LG_rescale}), in terms of the replica order parameter 
$\Omega_{\hat k}$, and replace $\Omega_{\hat k}$ by its expression 
in terms of the 
$(n+1)^{\rm th}$ moment of ${\cal N}\big(\{\rho_{\bf k}\}\big)$.
Thus, we arrive at the replica Landau free energy: 
\eqbreak
\begin{eqnarray}
nd {\cal F}_{n} & = & \epsilon -2 +
(3-\epsilon) \int 
{\cal D} \rho_1 \,{\cal N}\big(\{ \rho_{1,{\bf k}} \} \big)\,
{\cal D} \rho_2 \,{\cal N}\big(\{ \rho_{2,{\bf k}} \} \big)\,
\left(\sum\nolimits_{\bf k} \rho_{1,{\bf k}}\,\rho_{2,-{\bf k}}\right)^{n+1} 
\nonumber \\&& \qquad 
+\frac{1}{2}(n+1)\int 
{\cal D} \rho_1 \,{\cal N}\big(\{\rho_{1,{\bf k}}\}\big)\, 
{\cal D} \rho_2 \,{\cal N}\big(\{\rho_{2,{\bf k}}\}\big)\,
\left(\sum\nolimits_{\bf k}{k}^2\,
\rho_{1,{\bf k}}\, 
\rho_{2,-{\bf k}}\right)\,
\left(\sum\nolimits_{\bf k}\rho_{1, {\bf k}}\,
                  \rho_{2,-{\bf k}}\right)^{n} 
\nonumber\\&&\qquad
-\int 
{\cal D} \rho_1 \,{\cal N}\big(\{\rho_{1,{\bf k}}\}\big)\,
{\cal D} \rho_2 \,{\cal N}\big(\{\rho_{2,{\bf k}}\}\big)\, 
{\cal D} \rho_3 \,{\cal N}\big(\{\rho_{3,{\bf k}}\}\big)\,
\left(\sum\nolimits_{{\bf k}_1,{\bf k}_2} 
\rho_{1, {\bf k}_1}  
\rho_{2, {\bf k}_2} 
\rho_{3,-{\bf k}_1-{\bf k}_2} 
\right)^{n+1}.
\label{EQ:LG(n+1)_sdt_general}
\end{eqnarray}In 
order to obtain the desired (replica-independent) free energy we take
the limit $n\rightarrow 0$ of Eq.~(\ref{EQ:LG(n+1)_sdt_general}):
\begin{eqnarray}
d{\cal F}
&=&
d\,\lim_{n\to 0}{\cal F}_{n}=
(3-\epsilon) \int 
{\cal D} \rho_1 \,{\cal N}\big(\{\rho_{1, {\bf k}}\}\big)\, 
{\cal D} \rho_2 \,{\cal N}\big(\{\rho_{2, {\bf k}}\}\big)\,
\left(\sum\nolimits_{\bf k}\rho_{1,{\bf k}}\,\rho_{2,-{\bf k}}\right)\,
\ln\left(\sum\nolimits_{\bf k}\rho_{1,{\bf k}}\,\rho_{2,-{\bf k}}\right) 
\nonumber\\&&\qquad 
+\frac{1}{2}\int
{\cal D}\rho_1 \,{\cal N}\big(\{ \rho_{1,{\bf k}}\}\big)\, 
{\cal D}\rho_2 \,{\cal N}\big(\{ \rho_{2,{\bf k}}\}\big)\,
\left(\sum\nolimits_{\bf k}{k}^2 
\rho_{1,{\bf k}}\, 
\rho_{2,-{\bf k}}\right)\, 
\ln\left(\sum\nolimits_{\bf k}\rho_{1,{\bf k}}\,\rho_{2,-{\bf k}}\right)   
\nonumber \\&& \qquad
-\int 
{\cal D} \rho_1 \,{\cal N} \big( \{ \rho_{1,{\bf k}} \} \big)\, 
{\cal D} \rho_2 \,{\cal N} \big( \{ \rho_{2,{\bf k}} \} \big)\, 
{\cal D} \rho_3 \,{\cal N} \big( \{ \rho_{3,{\bf k}} \} \big)\,
\left(
\sum\nolimits_{{\bf k}_1,{\bf k}_2} 
\rho_{1, {\bf k}_1} \, 
\rho_{2, {\bf k}_2} \,
\rho_{3,-{\bf k}_1-{\bf k}_2}
\right)
\nonumber \\&& \qquad\qquad\qquad\qquad\times
\ln\left(\sum\nolimits_{{\bf k}_1,{\bf k}_2} 
\rho_{1,{\bf k}_1}\, 
\rho_{2,{\bf k}_2}\, 
\rho_{3,-{\bf k}_1-{\bf k}_2}\right).
\label{EQ:LG_sdt_general}
\end{eqnarray}In 
deriving the above free energy we have employed the 
physical fact that the average particle-density is uniform.  In other 
words, the replica order parameter is zero if all but one of the 
replicated wave vectors is nonzero which, translated in the language 
of the distribution of static density fluctuations, means that the 
first moment of the static density distribution equals 
$\delta_{{\bf k},{\bf 0}}$.  It is worth noting that, within this 
formalism, the replica limit can already be taken at the level of the free 
energy, prior to the hypothesizing of an explicit form for the 
stationary value of the order parameter.  On the one hand, this 
is attractive, as it leads to a Landau theory in which replicas play 
no role.  On the other hand, the approach is, at present, restricted 
to replica-symmetric states. 

We now construct the self-consistency condition that follows from the
stationarity of the replica-independent free energy. We 
then proceed to solve the resulting functional equation exactly, by
hypothesizing a solution having precisely the same physical content as
the exact solution of the replica self-consistency condition discussed
in Sec.~\ref{SEC:MFT_solve}.  

To construct the self-consistency condition for 
${\cal N}\big(\{\rho_{\bf k}\}\big)$ it is useful to make two observations. 
First, ${\cal N}\big(\{\rho_{\bf k}\}\big)$ obeys the 
normalization condition~(\ref{EQ:N_norm}). 
This introduces a constraint on the variations of 
${\cal N}\big(\{\rho_{\bf k}\}\big)$ which is readily accounted for 
via a Lagrange's method of undetermined multipliers. 
Second, as mentioned above, not all the variables $\{\rho_{\bf k}\}$ 
are independent: we have the relations 
$\rho_{\bf 0}=1$ and 
${\rho_{-{\bf k}}}= {\rho_{\bf k}^{\ast}}$. 
In principle, one could proceed by defining a new 
distribution that only depends on the independent elements of 
$\{\rho_{\bf k}\}$, 
and re-express the free energy in terms of this new
distribution.  However, for convenience we will retain 
${\cal N}\big(\{\rho_{\bf k}\}\big)$ as the basic quantity to be varied, 
and bear in mind the fact that $\rho_{\bf 0}=1$ and 
${\rho_{-{\bf k}}}={\rho_{\bf k}^{\ast}}$. 
By performing the constrained variation of 
$d{\cal F}$ with respect to the functional, 
${\cal N}\big(\{\rho_{\bf k}\}\big)$ 
\begin{equation}
0=\frac{\delta}{\delta {\cal N}\big(\{\rho_{\bf k}\}\big)} 
\left({\cal F}+
\lambda\int{\cal D}\rho_1\,{\cal N}\big(\{\rho_{1,{\bf k}}\}\big)\right), 
\end{equation}
where $\lambda$ is the undetermined multiplier, we obtain 
the self-consistency condition obeyed by 
${\cal N}\big(\{\rho_{\bf k}\}\big)$: 
\begin{eqnarray}
0&=&\lambda\,d
+ 2(3-\epsilon)
\int{\cal D}\rho_1\,
{\cal N}\big(\{\rho_{1,{\bf k}}\}\big)\, 
\left(\sum\nolimits_{\bf k}
\rho_{\bf k}\,
\rho_{1,-{\bf k}}\right)\, 
\ln\left(\sum\nolimits_{\bf k}
\rho_{\bf k}\,
\rho_{1,-{\bf k}}\right)
\nonumber\\&&\qquad
+\int{\cal D}\rho_1\,
{\cal N}\big(\{\rho_{1,{\bf k}}\}\big)\, 
\left(\sum\nolimits_{\bf k} {k}^2 
\rho_{\bf k}\,\rho_{1,-{\bf k}}\right)\, 
\ln\left(\sum\nolimits_{\bf k}
\rho_{\bf k}
\rho_{1,-{\bf k}}\right)
\nonumber\\&&\qquad
-3\int
{\cal D}\rho_1\,{\cal N}\big(\{\rho_{1,{\bf k}}\}\big)\, 
{\cal D}\rho_2\,{\cal N}\big(\{\rho_{2,{\bf k}}\}\big)\,
\left(\sum\nolimits_{{\bf k},{\bf k}^{\prime}} 
\rho_{\bf k}\,
\rho_{1,{\bf k}^{\prime}}\,
\rho_{2,-{\bf k}-{\bf k}^{\prime}}\right)\, 
\ln\left(\sum\nolimits_{{\bf k},{\bf k}^{\prime}}
\rho_{\bf k}\,
\rho_{1,{\bf k}^{\prime}}\,
\rho_{2,-{\bf k}-{\bf k}^{\prime}}\right).
\label{EQ:LG_sdt_sce}
\end{eqnarray}

To solve this self-consistency condition for 
${\cal N}\big(\{\rho_{\bf k}\}\big)$ we import our experience with the 
replica approach, thereby constructing the normalized hypothesis 
\begin{equation}
{\cal N}\big(\{\rho_{\bf k}\}\big)=
(1-q)\,\delta_{\rm c}(\rho_{\bf 0}-1)
\prod_{{\bf k}\neq{\bf 0}}\delta_{\rm c}(\rho_{\bf k}) 
+q\int\frac{d{\bf c}}{V}
\int_{0}^{\infty}d\tau\,p(\tau)\,
\prod_{\bf k}\delta_{\rm c}
(\rho_{\bf k}-{\rm e}^{i{\bf c}\cdot{\bf k}-k^2/2\tau}), 
\label{EQ:LG_sdt_ansatz}
\end{equation}
in which $q$ (which satisfies $0\le q\le 1$) is the localized 
fraction and $p(\tau)$ (which is regular and normalized to unity) 
is the distribution of localization lengths of localized particles. 
It is straightforward to show that by taking the $(n+1)^{\rm th}$ 
moment of ${\cal N}\big(\{\rho_{\bf k}\}\big)$ we recover the 
self-consistent form of the replica order parameter, 
Eq.~(\ref{EQ:ord_par_hyp}).

By inserting the hypothesis~(\ref{EQ:LG_sdt_ansatz}) into
Eq.~(\ref{EQ:LG_sdt_sce}), making the replacement 
$\rho_{\bf 0}\rightarrow 1$, and performing some algebra, the 
self-consistency condition takes the form 
\begin{eqnarray}
0&=& 
\int\frac{d{\bf c}}{V} 
\int_{0}^{\infty} d\tau\,
\left(1+
\sum\nolimits_{{\bf k}\neq{\bf 0}}
\rho_{\bf k}\, 
{\rm e}^{-i{\bf c}\cdot{\bf k}-k^2/ 2\tau}
\right)\, 
\ln\left(1+
\sum\nolimits_{{\bf k}\neq{\bf 0}}
\rho_{\bf k}\, 
{\rm e}^{-i{\bf c}\cdot{\bf k}-k^2/ 2\tau}
\right) 
\nonumber\\&&\qquad\qquad\qquad\times
\left\{ 
2q(-\epsilon +3q )p(\tau)
-q\frac{d}{d\tau}(2\tau^2p(\tau)) 
-3q^2\int_{0}^{\tau}d\tau_1\,p(\tau_1)\,p(\tau-\tau_1) 
\right\} 
\nonumber\\&&\qquad\quad
-\frac{3}{2}dq^2\int_{0}^{\infty} 
d\tau_1\,p(\tau_1)\,d\tau_2\,p(\tau_2)\, 
\ln\left\{V^{2/d}{\tau_1\tau_2}/{2\pi{\rm e}(\tau_1 +\tau_2)}\right\}
+\lambda\,d, 
\label{EQ:sdt_eqn_w/ansatz}
\end{eqnarray}in 
terms of the undetermined multiplier $\lambda$.
To determine $\lambda$ we insert the choice 
$\rho_{\bf k}=\delta_{{\bf k},{\bf 0}}$, which yields 
\begin{equation}
\lambda=\frac{3}{2}q^2
\int_{0}^{\infty}d\tau_1\,p(\tau_1)\,d\tau_2\,p(\tau_2)\,
\ln\left\{
V^{2/d}{\tau_1\tau_2}/{2\pi{\rm e}(\tau_1 +\tau_2)}\right\}. 
\label{EQ:lagrangian_multiplier}
\end{equation}
By using this result to eliminate $\lambda$ from the 
self-consistency condition, and observing that this condition 
must be satisfied for arbitrary $\{\rho_{\bf k}\}$, we arrive 
at a condition on $q$ and $p(\tau)$, viz., 
\begin{equation}
0=2q\,(-\epsilon +3q )\,p(\tau)
-q\frac{d}{d\tau}\Big(2\tau^2p(\tau)\Big) 
-3q^2\int_{0}^{\tau}d\tau_1\,p(\tau_1)\,p(\tau-\tau_1).
\label{EQ:old_one}
\end{equation} 
\eqresume
We integrate this equation over all values of $\tau$ and use the 
normalization condition on $p(\tau)$ to arrive at the same equation 
relating $q$ and $\epsilon$ as was found in Eq.~(\ref{EQ:q_eqn})
of the previous section, the appropriate solution of which is given by 
$q=2\epsilon/3$, i.e., Eq.~(\ref{EQ:q_soln}). Finally, we use this 
result for $q$ to eliminate it from Eq.~(\ref{EQ:old_one}), thus arriving 
at the same self-consistency condition on $p(\tau)$ as was found in 
Eq.~(\ref{EQ:scp}) 
of the previous section.
Thus, we see that these conditions, one for $q$ and
one for $p(\tau)$, are precisely the same as those arrived at by the
replica method discussed in Sec.~\ref{SEC:MFT_solve}.
\section{Comparison with numerical simulations: Universality exhibited}
\label{SEC:simulations}
The purpose of the present section is to examine the conclusions of 
the Landau theory, especially those concerning universality and 
scaling, in the light of the extensive molecular dynamics simulations, 
performed by Barsky and Plischke~\cite{REF:SJB_MP_a,REF:SJB_MP_b}.  
These simulations address the amorphous solidification transition in 
the context of randomly cross-linked macromolecular systems, by 
using an off-lattice model of macromolecules interacting via a 
Lennard-Jones potential.  
 \begin{figure}[hbt]
 \epsfxsize=\columnwidth
 \centerline{\epsfbox{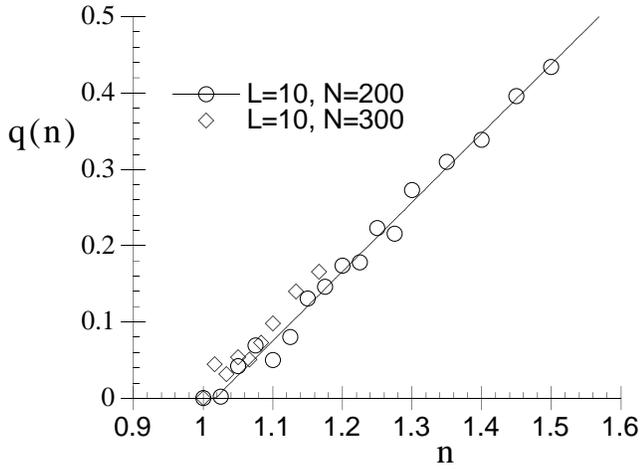}} 
 \vskip+0.5truecm
\caption{Localized fraction $q$ as a function of the number of 
cross-links per macromolecule $n$, as computed in molecular dynamics 
simulations by Barsky and Plischke (1997, unpublished).  
$L$ is the number of monomers in each macromolecule; 
$N$ is the number of macromolecules in the system.  
The straight line is a linear fit to the $N=200$ data.
Note the apparent existence of a continuous phase transition near $n=1$, 
as well as the apparent linear variation of $q$ with $n$, both features 
being consistent with the mean-field description.}
\label{FIG:gfr_sim}
\end{figure}
It should be emphasized that there are 
substantial differences between ingredients and calculational 
schemes used in the analytical and simulational approaches.  
In particular, the analytical approach:
(i)~invokes the replica technique; 
(ii)~retains interparticle interactions only to the extent that 
macroscopically inhomogeneous states are disfavored
(i.e., the one-replica sector remains stable at the transition); 
(iii)~neglects order-parameter fluctuations, its conclusions 
therefore being independent of the space-dimension; and 
(iv)~is solved via an Ansatz, which is not guaranteed to capture 
the optimal solution. 
\begin{figure}[hbt]
 \epsfxsize=\columnwidth
 \centerline{\epsfbox{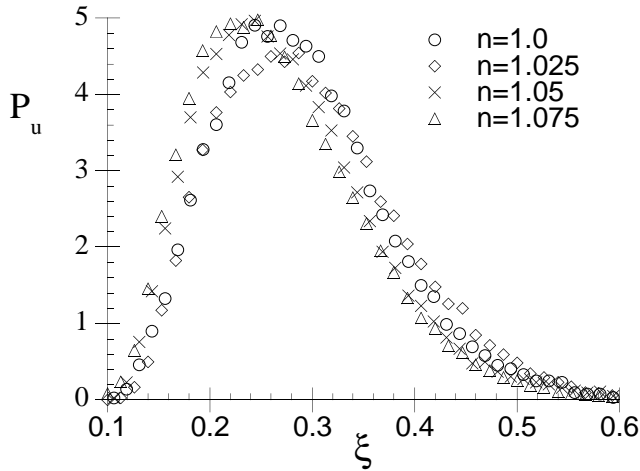}}
 \vskip+0.5truecm
\caption{Unscaled probability distribution $P_{\rm u}$ of localization 
lengths $\xi$ (in units of the linear system size), as computed in 
molecular dynamics simulations by Barsky and Plischke (1997, unpublished). 
In the simulations the number of segments per macromolecule is $10$; 
and the number of macromolecules is $200$.}
\label{FIG:dll_unscaled} 
\end{figure}
Nevertheless, and rather strikingly, the simulations yield an essentially 
identical picture for the transition to and properties of the amorphous 
solid state, inasmuch as they indicate that
(i)~there exists a (cross-link--density controlled) continuous phase
transition from a liquid state to an amorphous solid state;
(ii)~the critical cross-link density is very close to one cross-link per
macromolecule;
(iii)~$q$ varies linearly with the density of cross-links, at least in the
vicinity of this transition (see Fig.~\ref{FIG:gfr_sim}); 
(iv)~when scaled appropriately (i.e., by the mean localization length),
the simulation data for the distribution of localization lengths exhibit
very good collapse on to a universal scaling curve for the several
(near-critical) cross-link densities and macromolecule lengths considered
(see Figs.~\ref{FIG:dll_unscaled} and \ref{FIG:dll_scaled}); and
(v)~the form of this universal scaling curve appears to be in remarkably 
good agreement with the precise form of the analytical prediction for 
this distribution.  
\begin{figure}[hbt]
 \epsfxsize=\columnwidth
 \centerline{\epsfbox{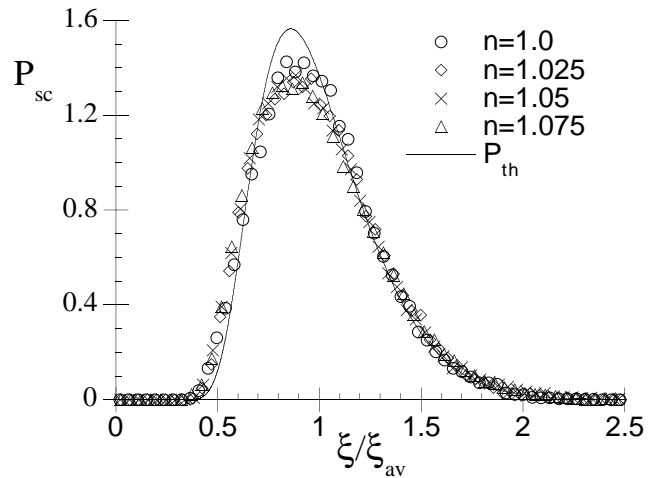}}
 \vskip+0.5truecm
\caption{Probability distribution (symbols) $P_{\rm s}$ of localization 
lengths $\xi$, scaled with the sample-average of the localization lengths 
$\xi_{\rm av}$, as computed in molecular dynamics simulations by Barsky
and Plischke (1997, unpublished). Note the apparent collapse of the data
on to a single universal scaling distribution, as well as the good 
quantitative agreement with the mean-field prediction for this 
distribution (solid line).  
In the simulation 
the number of segments per macromolecule is $10$; 
and the number of macromolecules is $200$.
The mean-field prediction for $P_{\rm sc}(\xi/\xi_{\rm av})$ is obtained 
from the universal scaling function $\pi(\theta)$ by 
$P_{\rm sc}(y)= (2s/y^3)\,\pi(s/y^2)$, where the constant 
$s\simeq 1.224$ is fixed by demanding that 
$\int_{0}^{\infty}dy\,y\,P_{\rm sc}(y)=1$.}
\label{FIG:dll_scaled} 
\end{figure}

It should not be surprising that by focusing on universal quantities, 
one finds agreement between the analytical and computational approaches.  
This indicates that the proposed Landau theory does indeed contain the 
essential ingredients necessary to describe the amorphous solidification 
transition. 

Let us now look more critically at the comparison between the results
of the simulation and the mean-field theory. With respect to the 
localized fraction, the Landau theory is only capable of showing the 
linearity of the dependence, near the transition, on the excess 
cross-link density, leaving undetermined the proportionality factor.  
The simulation results are consistent with this linear dependence, 
giving, in addition, the amplitude.  There are two facets to the 
universality of the distribution of localization lengths, as mentioned 
in Sec.~\ref{SEC:MFT_solve}.  First, that the distributions can, 
for different systems and different cross-link densities, be collapsed 
on to a universal scaling curve, is verified by the simulations, as 
pointed out above.  Second, the question of how the scaling parameter 
depends on the excess cross-link density is difficult to address in 
current simulations, because the dynamic range for the mean localization 
length accessible in them is limited, so that its predicted divergence 
at the transition is difficult to verify.
\section{Summary and concluding remarks}
\label{SEC:conclusions}
To summarize, we have proposed a replica Landau free energy for the 
amorphous solidification transition.  The theory is applicable to 
systems of extended objects undergoing thermal density fluctuations 
and subject to quenched random translationally-invariant constraints.  
The statistics of the quenched randomness are determined by the 
equilibrium density fluctuations of the unconstrained system.  We have 
shown that there is, generically, a continuous phase transition between 
a liquid and an amorphous solid state, as a function of the density of 
random constraints.  Both states are described by exact stationary points 
of this replica free energy, and are replica symmetric and macroscopically 
translationally invariant.  They differ, however, in that the liquid is 
translationally invariant at the microscopic level, whereas the amorphous 
solid breaks this symmetry. 

We have also shown how all these results may be recovered using an 
alternative formulation in which we focus less on the replica order 
parameter and more on the distribution of random static density 
fluctuations.  In particular, we construct a representation of the 
free energy in terms of this distribution, and solve the resulting 
stationarity condition. 

Lastly, we have examined our results in the light of the extensive 
molecular dynamics simulations of randomly cross-linked macromolecular 
systems, due to Barsky and Plischke.  Not only do these simulations 
support the general theoretical scenario of the vulcanization transition, 
but also they confirm the detailed analytical results for universal 
quantities, including the localized fraction exponent and the 
distribution of scaled localization lengths. 

The ultimate origin of universality is not hard to understand, despite 
the apparent intricacy of the order parameter associated with the 
amorphous solidification transition.  
As we saw in Secs.~\ref{SEC:MFT_replicas} and \ref{SEC:MFT_solve}, there 
are two small emergent physical quantities, the fraction of localized 
particles and the characteristic inverse square localization length of 
localized particles.  The smallness of the localized fraction validates 
the truncation of the expansion of the free energy in powers of the 
order parameter.  The smallness of the characteristic inverse square 
localization length leads to a very simple dependence, 
$\sum_{\alpha=0}^{n}\vert{\bf k}^{\alpha}\vert^{2}$, on the $n+1$ 
independent wave vectors of the replica theory, well beyond the permutation 
invariance demanded by symmetry considerations alone. As a result, 
near the transition, the amorphous solid state is characterizable in 
terms of a single universal function of a single variable, along 
with the localized fraction. 

Although throughout this Paper we have borne in mind the example of
randomly crosslinked macromolecular systems, the circle of ideas is by
no means restricted to such systems.  To encompass other systems 
possessing externally-induced quenched random constraints, such as
networks formed by the permanent random covalent bonding of atoms or
small molecules (e.g., silica gels), requires essentially no further
conceptual ingredients (and only modest further technical ones)
\cite{REF:GZ_glass}.

One may also envisage applications to the glass transition.  Although it
is generally presumed that externally-induced quenched random variables
are not relevant for describing the glass transition, it is tempting to
view the freezing-out of some degrees of freedom as the crucial
consequence of the temperature-quench, with a form of quenched disorder
thereby being developed spontaneously.  The approach presented in the
present Paper becomes of relevance to the glass transition if one
accepts this view of the temperature-quench, and thus models the
nonequilibrium state of the quenched liquid by the equilibrium state of
a system in which some fraction of covalent bonds has been rendered
permanent (the deeper the quench the larger the
fraction)~\cite{REF:sgrf}.  This strategy, viz., the approximating of
pure systems by ones with \lq\lq self-induced\rq\rq\ quenched disorder,
has also been invoked in very interesting work on the Bernasconi model
for binary sequences of low autocorrelation~\cite{REF:siqd}.
Interesting connections are also apparent with recent
effective-potential approaches to glassy magnetic systems, in which one
retains in the partition function only those configurations that lie
close to the equilibrium state reached at the glass transition
temperature~\cite{REF:franz}.

\noindent
{\it Acknowledgments\/} 
We thank Nigel Goldenfeld and David Hertzog for useful discussions.  
We gratefully acknowledge support from the U.S.~National Science 
Foundation through grants DMR94-24511 (WP, PG) and DMR91-57018 (PG), 
from a Graduate Fellowship at the University of Illinois at
Urbana-Champaign (HC), from NATO through CRG~94090 (PG, AZ), and 
from the DFG through SFB 345 (AZ). 
Michael Plischke has generously provided us with unpublished results 
from his extensive computational studies of vulcanized macromolecular 
networks, performed in collaboration with Sandra J.~Barsky, and has 
allowed us to report some of this work here.  We are most
grateful to him for this, and for his continued enthusiasm for the
subject-matter.  

\end{multicols}

\begin{references}
\bibitem{notation}
As is conventional in statistical mechanics, we use the same notation,
$\Omega_{\hat{k}}$, for the equilibrium value of the order parameter
and for the fluctuating variable (of which it is the expectation value
and which features in the Landau free energy). Which meaning is implied
can be readily ascertained from the context.
\bibitem{REF:sgrf}
This strategy has been discussed in greater depth in~\cite{REF:apy}; 
it has been implemented in Ref.~\cite{REF:GZ_glass}.
\end{references}
\end{document}